\documentclass[final,twocolumn,12pt]{elsarticle}
\usepackage[margin=1.5cm,includefoot]{geometry}

\usepackage{graphicx}
\usepackage{amssymb}
\usepackage{lineno}
\usepackage{mathtools}
\usepackage{appendix}
\usepackage[colorlinks]{hyperref}
\usepackage[nameinlink,capitalise]{cleveref}
\Crefname{figure}{Figure}{Figures}
\crefname{figure}{Fig.}{Figs.}
\crefname{appsec}{Appendix}{Appendices}

\usepackage{hyperref}
\usepackage{mathrsfs}
\usepackage{url}
\usepackage{booktabs}
\usepackage{caption}
\usepackage[para,flushleft]{threeparttable}
\usepackage{bbm}
\usepackage[separate-uncertainty=true]{siunitx}
\usepackage{subcaption}
\usepackage{capt-of}
\usepackage{multirow} 
\usepackage{dcolumn}
\usepackage{comment}
\usepackage{tikz}
\usetikzlibrary{positioning}
\usetikzlibrary{graphs}

\newcolumntype{d}[1]{D{.}{.}{#1}}

\makeatletter
\def\ps@pprintTitle{%
  \let\@oddhead\@empty
  \let\@evenhead\@empty
  \def\@oddfoot{\reset@font\hfil\thepage\hfil}
  \let\@evenfoot\@oddfoot
}
\makeatother

\creflabelformat{equation}{#2\textup{#1}#3}

\usepackage{stackengine}
\stackMath
\newcommand\tsup[2][2]{%
 \def\useanchorwidth{T}%
 \ifnum#1>1%
  \stackon[-1.3ex]{\tsup[\numexpr#1-1\relax]{#2}}{\mathchar"0365}%
 \else%
  \stackon[-1ex]{#2}{\mathchar"0365}%
 \fi%
}

\patchcmd{\emailauthor}{(#2)}{(C.W. Adair).}{}{}
\biboptions{sort&compress}

\begin{document}

\begin{frontmatter}

\title{A Decision Transformer Approach to Grain Boundary Network Optimization}

\author[myu]{Christopher~W.~Adair\corref{cor1}}
\ead{cwa367@byu.edu}
\author[myu]{Oliver~K.~Johnson}

\address[myu]{Department of Mechanical Engineering, Brigham Young University, Provo, UT 84602, USA}

\cortext[cor1]{Corresponding author.}

\begin{abstract}

As microstructure property models improve, additional information from crystallographic degrees of freedom and grain boundary networks (GBNs) can be included in microstructure design problems. However, the high dimensional nature of including this information precludes the use of many common optimization approaches and requires less efficient methods to generate quality designs. Previous work demonstrated that human-in-the-loop optimization, instantiated as a video game, achieved high-quality, efficient solutions to these design problems. However, such data is expensive to obtain. In the present work, we show how a Decision Transformer machine learning (ML) model can be used to learn from the optimization trajectories generated by human players, and subsequently solve materials design problems. We compare the ML optimization trajectories against players and a common global optimization algorithm: simulated annealing (SA). We find that the ML model exhibits a validation accuracy of $84\%$ against player decisions, and achieves solutions of comparable quality to SA (92\%), but does so using three orders of magnitude fewer iterations. We find that the ML model generalizes in important and surprising ways, including the ability to train using a simple constitutive structure-property model and then solve microstructure design problems for a different, higher-fidelity, constitutive structure-property model without any retraining. These results demonstrate the potential of Decision Transformer models for the solution of materials design problems.

\end{abstract}

\begin{keyword}
Grain Boundary \sep Grain Boundary Networks \sep Machine Learning \sep Decision Transformer \sep Human-Computer Interaction
\end{keyword}
\end{frontmatter}

\section{Introduction}
\label{S:Intro}

Grain boundary networks (GBNs) are high-dimensional structural features in microstructures that have been useful in modelling structure-property linkages \cite{Liu2023,Sayet2023,Bechtle2009,Snow2021,Johnson2018Spectral,Feng2022,Critchfield2020,Kamaya2009}. These models enable the simulation of grain boundary engineering, or the design of materials through manipulation of the microstructure \cite{Sergueeva2005,Snow2021,Bechtle2009,Tsurekawa1999,Feng2022}. Successes in grain boundary engineering have enabled materials to be designed with improved diffusivity \cite{Zheng2020,Bechtle2009,Tcherdyntsev2023,Hao2023,Oudriss2012}, corrosion resistance \cite{Liu2023,Feng2022,Zheng2020}, heat transfer \cite{Kontis2019,Meng2017,Guo2020Remarkably}, and mechanical properties \cite{Tsurekawa1999,Fast2008,Zhang2022}.

However, the dimensionality of the GBN state space is high: $3n_{GB}$ degrees of freedom for orientations and $2n_{GB}$ degrees of freedom for the boundary normals, where $n_{GB}$ is the number of grain boundaries, not including the additional degrees of freedom from GB connectivity \cite{Krakow2017,Homer2015}. Therefore, common design optimization methods, such as gradient ascent, are at risk of finding local, rather than global, maxima \cite{Lecchini-Visintini2009Simulated,Adair2022}.

Global optimization methods such as simulated annealing (SA) can overcome this weakness, but come with a computational efficiency trade-off \cite{Lecchini-Visintini2009Simulated}. Research into other high-dimensional systems such as quantum computing and protein folding found that human input in the form of a video game can act as a simplifying heuristic on their respective spaces, and as a result, can solve optimization problems in high-dimensional spaces more effectively than stochastic methods \cite{Quinn2011,Hasseriis2020,Khatib2010Crystal}. Previous work on GBNs specifically found that formulating a GBN design problem as a video game led to human players generating both better and more efficient optimization pathways than stochastic methods such as SA \cite{Adair2022}.

However, obtaining human inputs in this manner is expensive in time, development, and distribution costs. It is also difficult to use high-fidelity, but computationally expensive, material models within the constraints of video game design. Therefore, learning and/or replicating generalizable human optimization strategies could provide the benefits of human decision-making, while addressing the additional costs and limitations.

Machine learning (ML) is the obvious approach for this because of its ability to represent very high-dimensional information, generate optimization pathways, and learn from human inputs \cite{Zhai2021,Lee2022,Vaswani2017,Chen}. 

Transformer-based models have received much attention recently through GPT style implementations and classification tasks \cite{Vaswani2017,Chefer2021,Zhai2021,Lee2022}. Transformers function by learning the attention, or correlations, given to specific inputs, essentially creating a dictionary lookup of input to expected output \cite{Vaswani2017}. The goal of this representation is a more generalizable solution that can handle variable inputs and sets of states. This dictionary can be visualized to show how much certain states, inputs, actions, or words correlate with each other, giving a more accessible interpretation of the high-dimensional information \cite{Vaswani2017,Achache,Chefer2021}.

In this work, we generate a transformer-based ML model, and give a comparison of the solution quality and the solution efficiency of the players, the ML model, and a popular traditional global optimization algorithm (SA). We comment on the generalizability of the ML model. In particular, we explore generalization to a higher fidelity structure-property model without any additional training. 

\section{Background}

At the mesoscopic scale, grain boundaries are often defined by average grain misorientations and plane normals \cite{Krakow2017,Homer2015}. These structural definitions have been found to correlate with multiple material properties such as diffusivity \cite{Zheng2020,Bechtle2009,Tcherdyntsev2023,Hao2023,Oudriss2012}, corrosion \cite{Liu2023,Feng2022,Zheng2020}, heat transfer \cite{Kontis2019,Meng2017,Guo2020Remarkably}, mechanical properties \cite{Tsurekawa1999,Fast2008,Zhang2022}, and others \cite{Sergueeva2005,Zheng2020}. By informing material design from these discoveries, multiple advancements have been made in corrosion resistance \cite{Liu2023,Feng2022}, battery life extension \cite{Fu2023,Cheng2019Realizing}, desired diffusion effects \cite{Swiler1997,Hao2023}, hydrogen embrittlement \cite{Bechtle2009,Oudriss2012}, and others \cite{Zheng2020,Meng2017,Sergueeva2005,Zhang2022,Guo2020Remarkably,Kontis2019}. Generally, these successes have been achieved by implementing methods of material synthesis that enhance the \textit{statistical population} of targeted types of GBs \cite{Fu2023,Bechtle2009,Meng2017}.

Connectivity and other long-range effects of GBNs can also affect properties such as transport and fracture behavior \cite{Johnson2018Spectral,Adair2024, Fu2023,Liu2023}. Structure-property models that take these effects into account can be used to study and obtain optimal processes, states, and GB character that produce desired material performance, though design exploration of this space is necessarily complex due to the high number of interconnected degrees of freedom \cite{Homer2015,Sergueeva2005,Critchfield2020}. If exploration of this GBN design space can be done efficiently, then similar advances can be made to material design that can target \textit{spatial configurations} of boundaries in addition to their \textit{statistical populations}.

However, the configuration space of GBNs defined by the grain orientations and grain boundary plane normals becomes prohibitively large for conventional optimization or searching strategies such as gradient ascent, as well as having the increased risk of finding local, rather than global, maxima \cite{Lecchini-Visintini2009Simulated,Adair2022}. This is of particular concern for GB structure-property models, which possess sharp cusps due to crystallographic symmetries \cite{Bulatov2014}.

\begin{figure*}
    \centering
    \includegraphics{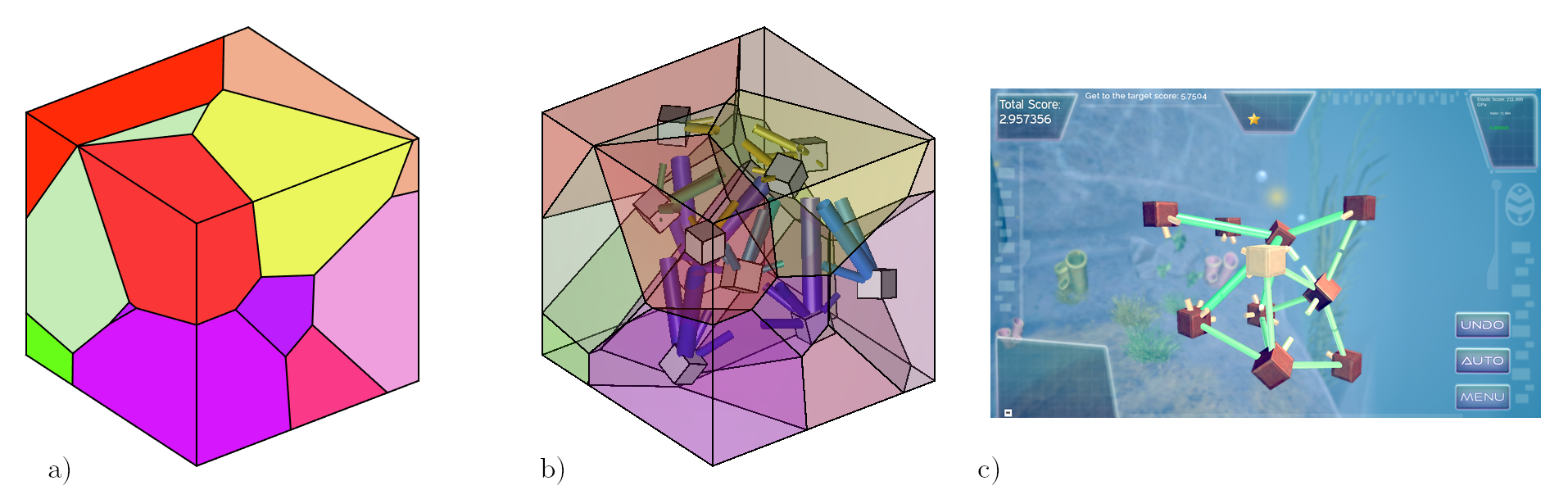}
    \caption{A visual description of how the GBN design problem was formulated in a video game context. (a) Shows an example microstructure as generated by Neper Polycrystal \cite{Quey2011Large-scale}. In (b) the grain centers are given a cube representing the lattice orientation of the respective grain, and the connections represent the magnitude of the corresponding GB property. In (c) the microstructure mesh is hidden, showing only the cubes and connections, and the actions and properties are given UI elements to communicate the design problem to players.}
    \label{F:MicrostructureToGame}
\end{figure*}

If the goal of a material design problem is searching for optimal structures within the design space, then most of the space likely consists of sub-optimal solutions that do not need extensive definition or attention, and, if skipped, could reduce the computations necessary to identify design candidates for more extensive testing \cite{Hasseriis2020,VonAhn2006,Krause2011Human}. Therefore, creating simplifying heuristics on GBN design optimization could enable design on not only statistical distributions, but boundary connectivity configurations as well (which requires a much larger design space).

One source for obtaining simplifying heuristics comes from human behavior, and has been studied extensively in human-computer interactions \cite{Quinn2011,VonAhn2008,Kawrykow2012Phylo:}. By formulating an optimization problem as a game, human inputs achieve the simplifications needed to optimize high-dimensional systems, like the GBN design problem \cite{Hasseriis2020,Adair2022,Kawrykow2012Phylo:,Khatib2010Crystal}. Previous work specifically for GBNs found that human inputs were capable of outperforming a global optimization algorithm on the design optimization of small GBNs, especially in decision efficiency \cite{Adair2022}.

There are multiple approaches to modeling GBNs for simulation. We use a graph representation, like previous work, that gives us a model of material performance based on the full connectivity and crystallographic information needed to describe GBNs \cite{Adair2022,Johnson2018Spectral}. The materials design video game from our previous work, Operation: Forge the Deep \cite{Adair2022a}, modeled the system as the dual of the GBN and allowed players to manipulate the crystallographic orientations of grains only (while keeping the geometry of the polycrystal fixed) \cite{Adair2022}. Each crystal orientation was visually represented as an orientable cube, while each grain boundary was represented as a connection between the pair of cubes (grain orientations) incident to the GB (see \cref{F:MicrostructureToGame}). Players were allowed to rotate each cube either manually, or with a local, greedy gradient ascent algorithm.

The length, girth, and color of each connection provided a real-time visual representation of the corresponding grain boundary property---diffusivity in this case---based on a constitutive model that took as inputs the GB plane and the orientations of the incident grains \cite{Adair2022,Johnson2018Spectral}. A green, fully touching, large diameter connection denoted an optimal property for that boundary in the game representation (see \cref{F:MicrostructureToGame}c).

The players' goal was to maximize the ``strength'' of these connections, thereby maximizing the material property of interest \cite{Adair2022}. The form this problem takes is then visibly recognizable as a variable size graph, where the nodes represent lattice orientations, and the edges represent grain boundary properties. This form of problem has been a popular subject for machine learning, aptly named graph-based learning \cite{Dwivedi2020,Mialon2021,Maskey2022,Park}.

Graph-based learning has been studied mainly in classification tasks, such as molecular identification, protein classification, and others \cite{Maskey2022,Park,Dwivedi2022}. However, the GBN design problem is not classification, but optimization as a sequence of decisions. Recent work has sought to tailor ML to sequence optimization through a new model called the transformer \cite{Vaswani2017}.

Transformer models are useful in this context because they allow for all kinds of inputs---provided there is an encoding available---and they are able to have flexible input sizes \cite{Vaswani2017}. The graph nature of problems is also included intrinsically through a position encoding. The position encoding then allows for learning on different configurations. However, transformers have tended to require large amounts of data for quality training \cite{Vaswani2017,Lee2022,Zhai2021}.

Decision Transformers are specific transformer models that model a time series of states, actions, and returns \cite{Lee2022,Chen}. Since the human decision-making process for the current problem closely matches this (i) state (the GBN configuration), (ii) action (grain selection and orientation change), and (iii) return structure (effective material property), we can apply this model well to our gathered player input data.

Other models exist, but other studies have already compared them against each other and found that decision transformers are the current best solution for highly variable state spaces of this structure \cite{Chen,Lee2022}.

We modify code created for the Multi-Game Decision Transformer \cite{Lee2022} and the Decision Transformer \cite{Chen} to generate this model. We expand their models to include multiple agents, which allowed for variable sized GBN simulations. We compare the ML performance against human player performance as well as SA, a common stochastic global optimization algorithm. 

We find that our modified Decision Transformer, trained on human player data, can effectively and efficiently optimize GBN properties. We detail the full model construction, validation, and evaluation and we explore ways in which the ML model generalizes.

\section{Methods}
\label{S:Methods}

We first describe the materials design problem, then we define the constitutive GB structure-property model, as well as the homogenization model to predict the effective property of the GBN as a whole. We then describe the three different optimization methods that we will compare for solution of the design problem: stochastic global optimization through SA, human inputs from a video game, and a Decision Transformer trained on human player data from the video game.

\subsection{The Microstructure Design Problem}
\label{S:M-Design}

In this study, we focus on the problem of designing a microstructure to maximize the rate of hydrogen diffusion along the GBN in polycrystalline Ni in a Type-C kinetic regime \cite{Harrison1961}. This design task is relevant for processes such as hydrogen separation during synthesis \cite{Dube2023}, where the permeation time of hydrogen through the nickel membrane depends partially on the diffusivity. Therefore, increases in membrane diffusivity can decrease the time required for hydrogen separation without requiring higher pressures. Additionally, nickel based materials can catalyze multiple hydrogen production reactions, leading to opportunities for combining production and separation operations \cite{Dube2023,Luczak2023}.

This particular design problem is given as an example, as our primary objective is the development of the computational method itself for application to GBN design problems generally. Indeed, the methods discussed here can easily be applied to arbitrary GBN design tasks, given suitable GB structure-property models.

The effective diffusivity of the GBN as a whole is calculated using a previously derived homogenization model which takes into account all crystallographic information and grain boundary connectivity information to predict the effective diffusivity ($D_{eff}$) of a 3D GBN. We summarize the key points here.

\subsubsection{The GB Structure-Property Model}
\label{S:M-ConstModel}

\begin{figure*}
    \centering
    \includegraphics{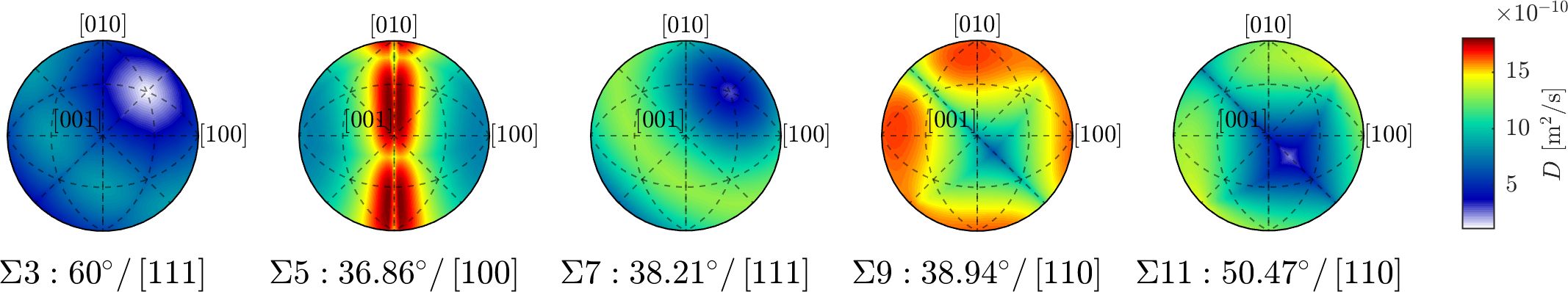}
    \caption{Borisov/BRK diffusivity model for H in Ni GBs. Subplots show the GB diffusivity as a function of the GB normal in the crystal reference frame (see annotated crystal directions) for several low-$\Sigma$ misorientations.}
    \label{F:BorisovBRKDiffusivityModel}
\end{figure*}

A required input for the homogenization model \cite{Johnson2018Spectral} is a constitutive GB structure-property model that takes the 5 crystallographic degrees of freedom of each GB as input (3 for misorientation + 2 for plane inclination), and returns the value of the GB property of interest (in this case, the diffusivity of H in Ni GBs). For the present work, we will employ the model from Page, et al. for GB diffusivity of hydrogen in nickel \cite{Page2021}. This model is comprised of (i) a refinement of the Borisov relation, which predicts GB diffusivity from GB energy \cite{borisov1964relation}, and (ii) the Bulatov, Reed, and Kumar (BRK) model for GB energy as a function of the 5 crystallographic degrees of freedom of the GB \cite{Bulatov2014}. This model was validated against molecular dynamics simulations of H diffusion in Ni GBs \cite{Page2021}. Advances in Scanning-TEM have also allowed for some experimental validation of the Borisov relation at low temperatures \cite{Schweizer2023}.

\Cref{F:BorisovBRKDiffusivityModel} shows the resulting model for H diffusivity in Ni GBs as a function of 5D GB character. The subplots show the GB diffusivity as a function of the GB plane normal (see annotated Miller indices) for several low-$\Sigma$ misorientations. As is apparent, the variation of GB diffusivity is a complex function of the 5D GB character.

The materials design video game that provided the training data presents real-time predictions of the effective diffusivity of the GB network (the ``score'' that the players see) as players manipulate the microstructure. This requires the use of a constitutive model that is fast to evaluate. Unfortunately, the Borisov/BRK diffusivity model just described, while realistic, is too computationally expensive to provide the needed real-time property calculations. Instead, the training data was collected using the following simple toy model for GB diffusivity:
\begin{equation}
    D=\beta\left(\frac{\theta}{10} +\left|N_x\right|+\left|N_y\right|+\left|N_z\right|\right)
    \label{E:Constitutive}
\end{equation}
where $\theta$ is the disorientation angle between the neighboring grains, $\mathbf{N} = \left[N_x,N_y,N_z\right]^\mathsf{T}$ is the GB plane unit normal in the lab frame, and $\beta = {10}^7$ is an arbitrary scaling parameter. We will refer to this as the ``Linear model''. The form of this model is similar to test functions used for GB properties in other studies \cite{Mohles2020,Naghibzadeh2024}. This simple model satisfies the computational performance requirements of the video game, while retaining dependence on some of the GB parameters and facilitating investigation of the high-dimensional GBN design optimization problem using the video game. This is purely a computational expedient. Nevertheless, we hypothesize that the ML model can still learn useful optimization strategies from the resulting training data. We then attempt to apply the ML model, in the context of the realistic Borisov/BRK constitutive model, without retraining. This would imply that the ML model could be trained using one (simple) constitutive GB model, and then generalize to make predictions for a different (higher-fidelity) constitutive GB model. Testing this hypothesis of generalizability is one of the objectives of the present work and will be discussed later.

\subsubsection{3D Microstructure Representation}
\label{S:M-3DMicro}

The GB normals required by the constitutive GB structure-property models are obtained from a surface mesh of triangular elements along the grain boundaries of a given microstructure for a finite volume (FV) method representation \cite{Adair2022}. Forty-six 3D microstructures were generated and meshed using Neper polycrystal \cite{Quey2011Large-scale} and given grain growth morphologies to better represent realistic samples. The generated microstructures spanned 10 different numbers of grains, ranging from 4 to 35. An example of a generated structure can be seen in \cref{F:MicrostructureToGame}.

To fully define a FV element, we set the grain boundary thickness at a constant \si{5 \angstrom}, while the remaining geometric parameters of the GB are determined from the local mesh geometry.

\subsubsection{Calculating Effective Diffusivity}
\label{S:M-Deff}

The effective diffusivity of H across the entire GBN ($D_{eff}$) is calculated through a mass flow PDE, where mesh vertices on the boundary are given either Dirichlet or Neumann boundary conditions. We follow the same boundary conditions as our previous work \cite{Adair2022}, where the faces perpendicular to the $x$-direction have Dirichelt conditions for the concentrations of $c_{source}=1\,\si{ \kg / \m^3}$ and $c_{sink}=0\,\si{\kg / \m^3}$ respectively, and all other directions have Neumann boundary conditions applied. For computational simplicity, all Dirchlet condition boundary vertex indices are combined into a single source ($v_{source}$) or sink ($v_{sink}$) index, respectively. This maintains the edge information of the mesh, but simplifies the PDE setup.

The Laplacian, $\boldsymbol{\mathcal{L}}$, of the GBN mesh is calculated according to:
\begin{equation}
  \mathcal{L}_{ij}=
  \begin{dcases}
     \sum_{i\sim m}\frac{D_{im}A_{im}}{L_{im}} & \text{if }i=j\\
     -\frac{D_{ij}A_{ij}}{L_{ij}} & \text{if }i\sim j\\
     0 & \text{otherwise}\\
  \end{dcases}
  \label{E:GraphLaplacian}
\end{equation}
where $D_{ij}$ is the diffusion coefficient assigned to the mesh edge connecting vertices $i$ and $j$ as calculated from the chosen GB constitutive model (the Linear model or the Borisov/BRK model), $A_{ij}$ is the element area, and $L_{ij}$ is the element length (see \cite{Adair2022} for detailed definitions).

Following the method described in \cite{Kurniawan2020}, we can compute $D_{eff}$ through the total mass flow to the sink from the source according to
\begin{equation}
    D_{eff}=\frac{l_s}{w_s t_s c_{source}}\left(-\boldsymbol{\mathcal{L}}_{[v_{sink},\cdot]}\mathbf{Q}^{-1} \mathbf{b}\right)
    \label{E:Deff}
\end{equation}
where $l_s$, $w_s$, and $t_s$ are the length, width and thickness of the material sample respectively, $\boldsymbol{\mathcal{L}}_{[v_{sink},\cdot]}$ denotes the $v_{sink}$-th row of $\boldsymbol{\mathcal{L}}$, $\mathbf{b} = c_{source}\mathbf{e}_{source} + c_{sink}\mathbf{e}_{sink}$, and $\mathbf{Q}$ is defined as
\begin{equation}
    \mathbf{Q}=\left[
    \begin{matrix}
        \boldsymbol{\mathcal{L}}_{[F,\cdot]}\\
        \mathbf{e}_{source}^\mathsf{T}\\
        \mathbf{e}_{sink}^\mathsf{T}
    \end{matrix}
    \right]
\end{equation}
where $F$ is the set of vertices excluding the source and sink (i.e. $\boldsymbol{\mathcal{L}}_{[F,\cdot]}$ is $\boldsymbol{\mathcal{L}}$ with the source and sink rows removed), with $\mathbf{e}_i$ denoting the vector whose $i$-th element is $1$ and all others are $0$. For a more comprehensive explanation of the derivation we refer the reader to \cite{Kurniawan2020}.

Given this homogenization relation, the objective of the present GBN design problem is to maximize $D_{eff}$, by changing the crystallographic orientations of grains in the polycrystal.

\subsection{Optimization through a Stochastic Method}
\label{S:M-SimAnneal}

The first method we consider for optimizing $D_{eff}$ is a stochastic global optimization algorithm called simulated annealing (SA) \cite{Lecchini-Visintini2009Simulated}. SA functions by taking random steps through the optimization landscape and either accepting a step that \emph{improves} the solution, or accepting a step that \emph{worsens} the solution with a given probability that decreases over time \cite{Lecchini-Visintini2009Simulated}. Accepting worse steps enables the solution to avoid local maxima better than simple gradient ascent, but comes at a computational cost. This is especially true for complex, expensive models such as the Borisov/BRK model.

For the GBN optimization, we use a Cauchy annealing schedule \cite{Dukkipati2004} where at each Monte Carlo step a grain is randomly selected, via uniform distribution, from the full GBN and assigned a new orientation, sampled uniformly from SO(3). We define the convergence criterion to be 1000 consecutive rejected steps. Five runs of SA using the Linear model were done for each of the evaluation microstructures, which will be defined subsequently. However, due to the complex and computationally expensive nature of the Borisov/BRK model, only one Borisov/BRK model run of SA for each evaluation structure was possible with the available computational resources.

\subsection{Optimization through a Video Game}
\label{S:M-VideoGame}

The second method we consider for optimizing $D_{eff}$ is through human inputs via a video game. The game, ``Operation: Forge the Deep'', is a 3D puzzle game where users manipulate cubes, representing grain orientations, to change the connections, representing individual GB properties, to optimize a score, representing the effective material property of the polycrystal \cite{Adair2022a}. A screen-shot of the video game is shown in \cref{F:MicrostructureToGame}. 

Players were allowed to perform one of the following actions to modify the orientation of a single grain at a time:
\begin{itemize}
    \item Apply a manual rotation
    \item Apply a locally optimal rotation via gradient ascent
    \item Undo the previous action
\end{itemize}
Additionally, certain scenarios (levels) in the game antagonistically introduce sporadic random rotations to grains, meant to throw off the player's thought process and increase the challenge.

The players' goal in each scenario presented to them (i.e. each level) was to reach a score corresponding to 90\% of the maximum $D_{eff}$ solution found for that GBN morphology using SA with the Linear model. GBN morphologies were repeated in separate scenarios with additional restrictions such as: limited number of actions, limited time to solve, limited grain selection, and combinations of these restrictions. These constraints were implemented in different game ``levels'' to encourage player engagement, but they also may encourage varied solution approaches, thereby diversifying the training set for the ML approach.

Previous studies on human inputs in this game environment were done in a laboratory setting on a small scale and with specific conditions of interest \cite{Adair2022}. However, the current work uses data collected from a public distribution of the game through the Steam storefront, in accordance with data privacy agreements and releases \cite{Adair2022a}. Additionally, this game was presented to K-12 students as part of a STEM outreach program that also contributed to data collection. In total, 879 optimization trajectories were collected from users, which consisted of between 5 and 810 decisions per trajectory.

These trajectories were stored as a sequence of decisions, where the action, resultant state of all grain orientations, and resultant score were stored in sequence.

\subsection{Optimization through a Decision Transformer}
\label{S:M-DecisionTran}

\begin{figure}[h!]
    \centering
    \includegraphics[]{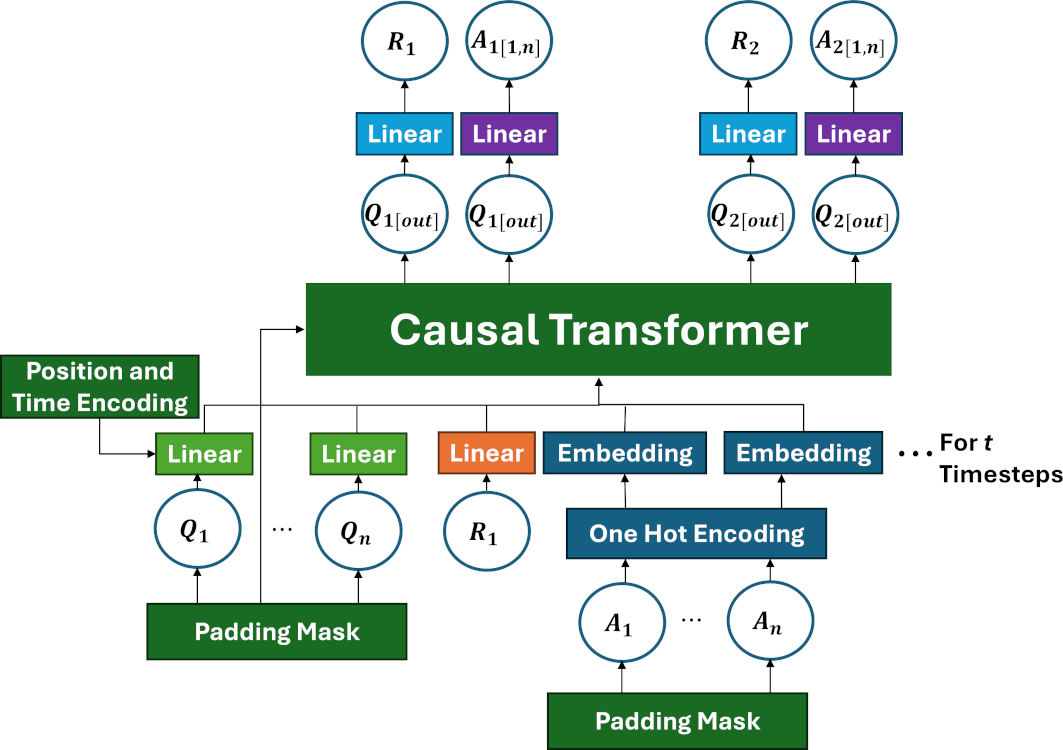}
    \caption{Block diagram of the GBN design Decision Transformer ML model.}
    \label{F:BlockDiagram}
\end{figure}

The third method we consider for optimizing $D_{eff}$ is creating, training, and evaluating a Decision Transformer that uses sequences of player decisions, called player trajectories. While we base the implementation of this Decision Transformer on the original \cite{Chen} and the Multi-Game Decision Transformer \cite{Lee2022}, there were substantial changes that needed to be made to apply the technique to the GBN design problem. An overview of the ML model can be seen in \cref{F:BlockDiagram}, which will be described in detail.

\subsubsection{State, Action, and Return Space}
\label{S:M-SAR Space}

To prepare player data for training of the Decision Transformer model, the state, action, and returns must be clearly defined and quantified. The state space observed by the players includes the set of grain orientations and the strength of the connections between them. The grain orientations were represented as a set of quaternions. The shape of the resulting state tensor is $[B \times t \times n \times 4]$, where $B$ is the batch size, $t$ is the number of time steps included, $n$ is the number of grains in a simulation, and $4$ represents the quaternion information (each quaternion is a 4D unit vector).

The return is simple to implement, as the game score (value of $D_{eff}$) represents the full return. However, normalizing the return requires a new definition. As the GBN design problem is a configurational optimization problem with many local maxima, it is unknown \emph{a priori} if a configuration exists in which all boundaries exhibit the maximum diffusivity allowed by the underlying GB constitutive model while also satisfying all physical (e.g. crystallographic) constraints.

Therefore, while the true maximum may be unknown, a standardized normalization can be calculated by artificially setting all boundaries in a given microstructure to the maximum value of the selected GB constitutive model, and then calculating the resulting value of $D_{eff}$ as an upper bound on the theoretical maximum. Each structure is normalized by its respective upper bound value, so all returns are scaled between 0 and 1. This encourages generalizability of training both between microstructure morphologies and possibly material models as well. The resulting input tensor is of the shape $[B \times t \times 1]$.

The action space representation differs from other Decision Transformers, since $n$ can have a variable size (different levels correspond to polycrystals with different numbers of grains), therefore, there is a variable size in the action space as well. The action space is set as a one-hot encoding consisting of 5 possible actions for each grain:
\begin{enumerate}
    \item Do nothing
    \item Manual rotation
    \item Local gradient ascent
    \item Undo
    \item Assign random orientation
\end{enumerate}

The resulting input tensor is of the shape $[B \times t \times n \times 5]$, with the same variable meanings as the state tensor. Although only one grain is changed during a given step, the ML model technically updates the state of \textit{all} grains at each step; thus, the ``Do nothing'' action is simply what is assigned to all grains that were not selected for the active action assigned at that step (actions 2-4 in the list above).

\subsubsection{The Input Layer}
\label{S:M-InputLayer}

Each input tensor is then projected to the hidden size, $h$, for the ML model. The state and return embeddings use a Linear layer each, resulting in tensor shapes of $[B \times t \times n \times h]$ and $[B \times t \times 1 \times h]$ respectively. The actions embedding uses an Embedding layer, which is useful for the sparse nature of the information, and results in a tensor shape of $[B \times t \times n \times h]$.

To account for the variable value of $n$, a mask is applied to the state and action tensors that prepends zeros to each tensor such that, regardless of the actual number of grains in a particular problem, the tensor will have a size of $n_{pad} = 50$ in the $n$ dimension \cite{Reddy}. This mask is saved so the ML model does not take into account these zeros during training or evaluation.

With the $n$ dimension having fixed size, the input tensors can be combined in a way that models the decision-making nature of the inputs. The inputs are interleaved according to $t$ so that the information has the order state, return, action, for each time step $t$. This follows the logic of previous Decision Transformers, stating that an expected return and selected action can be predicted from an observed state, and that future actions are influenced by previous decisions \cite{Lee2022,Chen}.
    
\subsubsection{The Position Encoding}
\label{S:M-PosEncode}

As constructed, the input tensors do not explicitly include information about the GB properties nor the order in which decisions were made. Both kinds of information can be added through a \emph{position encoding} \cite{Dwivedi2020,Park,Mialon2021}. Position encodings have been used especially in large language models as a method for passing relative position information of tokens to the ML model for training \cite{Vaswani2017,Zhai2021,Achache}. Graph-based learning has also developed similar methods for passing graph edge information alongside node information \cite{Dwivedi2020,Park,Mialon2021}.

For this model, we utilize the Laplacian Position Encoding (LPE) first proposed by Dwivedi, et al. for use in ML transformers \cite{Dwivedi2020}. LPE calculates the eigenvectors of the weighted Laplacian matrix, constructed from a given graph, embeds the values, and adds the layer element-wise to the graph node representation. 

In our model, the nodal information of the GBN dual graph consists of the state and action tensors, which each encode grain (node) data, and the GB (edge) information. This is calculated from the selected GB constitutive model for all grain neighbors after every decision step. A Laplacian can be calculated similar to \cref{E:GraphLaplacian} for the GBN puzzle, which takes the nodes as the grains (cubes) and edges as the GB properties (connections), which can then be used to calculate eigenvectors and eigenvalues corresponding to the current state of the microstructure. While the absolute values of GB properties are vastly different between the Linear and Borisov/BRK models, this step acts to normalize the relative property connections to comparable, normalized eigenvectors.

Since the number of eigenvectors is equal to the number of nodes (grains) in a graph, the most that can be calculated for the smallest simulation is 4. To keep the amount of information constant for training, we take the first 4 eigenvectors, sorted by increasing eigenvalue, of each state as the position encoding. After embedding through a Linear layer, the shape of the resulting tensor is $[B\times 1 \times n_{pad} \times h]$, which is calculated for every $t$ and added to the corresponding state and action tensor.

An additional position encoding is made for the time step associated with each decision. An encoding vector is made for each token (individual state, return, and action inputs), containing the time step in which the token was made. The shape of the tensor is $[B\times T]$, where $T$ is the total number of tokens (state, return, and action for all $t$) in an input, and the value stored is the timestep the token appeared in. A maximum value of 810 was set for timesteps, though training data rarely reached that many decisions. After embedding with an Embedding layer, the shape of the Time Encoding tensor is $[B\times T \times h]$. This is also added to all tokens in the input layer.

\subsubsection{The Transformer Layer}
\label{S:M-Transformer}

After inputs have been embedded and encodings added, the tensor is fed into a Causal Transformer \cite{Vaswani2017}. The purpose of the transformer layer is to create a ``soft'' dictionary, where the inputs and their correlations (the query), match a known state (the key), which further has a known solution (the value) \cite{Vaswani2017}. The ``soft'' nature of the lookup means that there does not need to be an exact match for each state (query and key) and output (value), but general trends and meanings are still captured \cite{Vaswani2017}.

A Causal Transformer further enforces the logic of sequential decisions by masking future tokens away from past tokens. This disallows tokens (inputs) from affecting training for future steps, while allowing future tokens to have ``memory'' of past tokens \cite{Vaswani2017,Lee2022}. The Causal Transformer consists of three major sections: the Block Causal Mask; the Query, Key, and Value layers; and the feed-forward layer.

\subsubsection{The Block Causal Mask}
\label{S:M-BlockCausal}

A traditional causal mask is a lower triangular matrix, which represents the sequential dependence of time series information, where the future tokens can remember the past, but the past tokens cannot know the future. The Block Causal Mask is an improvement to the information representation implemented in the Multi-Game Decision Transformer \cite{Lee2022} that allows concurrent, but separate, pieces of information to attend to each other, unlike a traditional causal mask. This is constructed by creating a lower triangular matrix of size $[B\times T \times h]$ and value of one, and then adding a ``block'' of $[N\times N]$ ones, where $N$ is the number of simultaneous observations, starting at each diagonal entry where the observation tokens begin \cite{Lee2022}.

In the Multi-Game Transformer, only the observations (sections of screen pixels) were given this consideration \cite{Lee2022}, while the GBN materials design game requires more. As all grains are considered their own agent, each grain has an action assigned during a time step, and should be allowed to ``see'' every grain take their action. Therefore, an additional $[N\times N]$ block is added for the actions (player decision) as well as the observations (grain orientations).

\subsubsection{The Query, Key, and Value Layers, and Feed-Forward}
\label{S:M-QKV}

All three Query, Key, and Value (QKV) layers are Linear layers of the same shape as the final input layer that are combined to form a transformer block \cite{Vaswani2017}. The input layer is fed through a Linear normalization layer, and then each of the three QKV layers in parallel. The equation for the QKV relation that defines the attention (attn) is then
\begin{equation}
    attn = \text{soft}(\text{mask}(Q \times K^\mathsf{T}/\sqrt{h}))V
    \label{E:Transformer}
\end{equation}
where ``soft'' is the softmax function, ``mask'' is the application of the block causal mask, $Q$ is the query layer, $K$ is the key layer, $V$ is the value layer, and $h$ is the hidden size. The cross product is only for the final two dimensions of each layer $[T\times h]$, as $B$ the batch variable is only for parallel computations. The result is then projected back into the correct dimensionality by a Linear layer.

The resulting outputs (``logits'' in ML terminology) are fed through a simple multi-layer perceptron (MLP) with a GeLU activation \cite{Lee2023}. During training there are also dropout layers added after the QKV block and the MLP block.

\subsubsection{Output Layer}
\label{S:M-OutputLayer}

The logits at this step have identical shape to the inputs, with interleaved state, return, and actions. The final layer is what creates the correlation between previous states and future returns and actions. Each set of logits corresponding to the original states are fed to a Linear layer, $[h \times 5]$. The resulting output is the predicted action. Unlike other ML models, since there is a one-to-one matching of states to actions we can use the entire state information to predict future actions, instead of a single output like previous models \cite{Lee2022,Vaswani2017}.

For the returns, however, only the final output of each state is fed to a Linear layer, $[h \times 1]$, to predict the next return. By using the state information in this way, the next action and return become a response to the observed state.

\subsubsection{Hyperparameters and Training}
\label{S:M-HyperAndTrain}

For training and evaluating the model we set the batch size $B=16$, hidden size $h=1024$, and use a time step window of $4$. This window sets the total tokens as $T=404$. For training we set the optimizer as AdamW \cite{Loshchilov2019}, the learning rate at $10^{-7}$, and the decay at $10^{-6}$. The low learning rate was chosen to avoid overfitting to the data. We use 1000 warmup steps and run each training for \num{10000} steps.

The training set contained 897 player trajectories consisting of between 5 and 810 decisions each. The window of 4 decisions were taken at random from a trajectory for each step of the training, following similar training methodology to \cite{Lee2022}.

A major benefit of this method is the ability to train on all data points, regardless of the quality of the training data. Since the transformer acts as a ``soft'' dictionary, all data contributes to knowing not only what are optimal optimization steps, but also what are non-optimal steps to be avoided.

The loss function which is minimized during training was set as the sum of the cross entropy of predicted vs. actual actions (due to one-hot encoding) and the Euclidean distance between predicted vs. actual returns. Accuracy was defined as the mean of the percent of correct actions chosen and the relative error between returns.

Validation of the model was calculated on a sequence basis during training, which was measured by the accuracy of the whole 4 time-step prediction.

As the training methodology samples the training data using random number generation, there will be some training stochasticity effects on model inference. Therefore, we will train and evaluate the ML model 10 times to evaluate the resulting variability.

\subsubsection{Evaluation}
\label{S:M-Eval}

The ML model was evaluated by holding 4 meshed microstructures, and the associated player trajectories, out of the training set. The microstructures had 10, 15, 25, and 30 grains respectively. Prediction was run for 810 decision steps so as to remain within the training data available. Chosen actions and total return (i.e. the game score, which was simply the value of $D_{eff}$) were saved for each time step.

At prediction time, the ML model only provides the action type to be applied and the selected grain to apply the action to. Algorithms are available from the video game implementation for actually applying 4 of the 5 actions to the microstructure. However, the manual rotation action must be generated separately at prediction time. Rather than training the ML model to predict exactly what manual rotation would be taken for a given step, we instead model it.

We model the manual movement of players as a local gradient ascent that maximized the properties of the connections to the currently selected grain. We justify this decision from observing player movement, where we observed that players tended to follow a rough gradient ascent in their decisions, though sometimes they overshot a local maximum. Therefore, we believe modeling the local movement as a simple gradient ascent on the local property to be an appropriate estimate of the manual motion.

The trained model is given the current state, and any previous states and actions up to the time-step window of 4, which gives the logits for the actions and the expected return. The action logits' output are the probability of choosing a given action out of the 5 available for each grain in the current state. A single action is chosen by maximum probability to be applied to the GBN simulation. The GBN orientations, boundary properties, and position encodings are then updated and used as the next state input.

Because not all player inputs used in training were optimal optimization trajectories (all data were used, not just good optimization trajectories), without additional information the predictive performance would be limited. However, as stated previously, a strength of the transformer model is the dictionary structure of the learning. 

Therefore, at prediction time we add a flat bias of 0.1 (10\% of maximum return) to the expected return output, and re-predict the same time step. This bias encourages the ML model to search for more optimal solutions among the training that can reach the biased score in a single step, attempting to emulate an expert (more optimal) player trajectory at prediction time \cite{Lee2022}.

The bias of 0.1 would then imply that, in the ideal scenario, the ML model would reach the maximum return in 10 decision steps. This will most likely not be the case, especially with larger grain numbers. However, the distribution of step counts for top performing players (95\textsuperscript{th} percentile of all players by max return) achieving their max return is heavily skewed towards 10 steps or less in the training set. Therefore, we feel the selected bias is appropriate for attempting to generate expert trajectories.

\subsection{Quantifying and Comparing Optimization Methods}
\label{S:M-Quant}

To compare the three optimization methods (SA, Player, ML) we use the same comparisons as previous work: solution quality and solution efficiency \cite{Adair2022}. Solution quality is given by the normalized Return (the achieved material property value, $D_{eff}$, divided by the hypothetical upper bound value, as described in \cref{S:M-SAR Space}). A value of 1 represents a perfect solution, while 0 represents the worst possible. This is the same as the return input to the ML model. 

For comparing the ML model against players and SA, we will consider both the best and median performing models across the 10 trained model replicates. The best ML model will be compared to the best player or the best SA, and the median performing ML model will be compared against the median player or the median SA. 

Each trained model is applied to each of the 4 evaluation microstructures. The performance of each model replicate is defined to be the \textit{smallest} maximum return across the set of 4 evaluation microstructures. That is, for a given model replicate, we consider the highest return achieved across time steps for each evaluation microstructure; the smallest of the 4 resulting values is defined to be the performance of that model replicate. This represents a worst-case measure of performance across all of the evaluation microstructures. This definition was chosen to enforce the idea that a trained model should perform well on all microstructures, not just a single one that skews the average high. Thus, the ``best'' and ``median'' ML models refer, respectively, to the replicates which exhibit the best and median performance using this definition.

To compare solution quality, we will use the following definition
\begin{equation}
\label{E:DeltaReturnDefinition}
    \Delta \text{Return}_{X} = \text{Return}_{ML} - \text{Return}_{X}
\end{equation}
where $X$ is either $Player$ or $SA$. Note that $\Delta \text{Return}_{X} > 0$ implies the ML model achieved a higher return and $\Delta \text{Return}_{X} < 0$ implies the other method achieved a higher return.

The solution efficiency is measured by the number of decisions taken to achieve a fixed solution quality for the first time \cite{Adair2022}, which we choose to be the smaller of the two maximum returns of the methods being compared, according to
\begin{equation}
\label{E:DeltaStepsDefinition}
    \Delta \text{Steps}_{X} = \text{Steps}_{X} - \text{Steps}_{ML}
\end{equation}
Note that the order of arguments in \cref{E:DeltaStepsDefinition} is reversed compared to \cref{E:DeltaReturnDefinition}. This is intentional so that positive values always indicate better ML performance (in terms of solution quality when $\Delta \text{Return}_{X} > 0$ or in terms of efficiency when $\Delta \text{Steps}_{X} > 0$).

We also test the generalizability of the ML model to other constitutive GB structure-property models that it was not trained on (the Borisov/BRK model). We will do this by repeating the evaluation on the 4 microstructures, but replacing the Linear model with the Borisov/BRK model during evaluation. No additional training nor changes to ML model inputs are done to accomplish this change.

For each of the 4 evaluation microstructures, we will compare the difference between the solution quality and solution efficiency of the ML model vs. the players, and the ML model vs. SA. However, since the Borisov/BRK model does not have any player trajectories, we will only compare the ML model vs. SA.

\section{Results}

\subsection{Training and Validation}
\begin{figure}[h]
    \centering
    \includegraphics[]{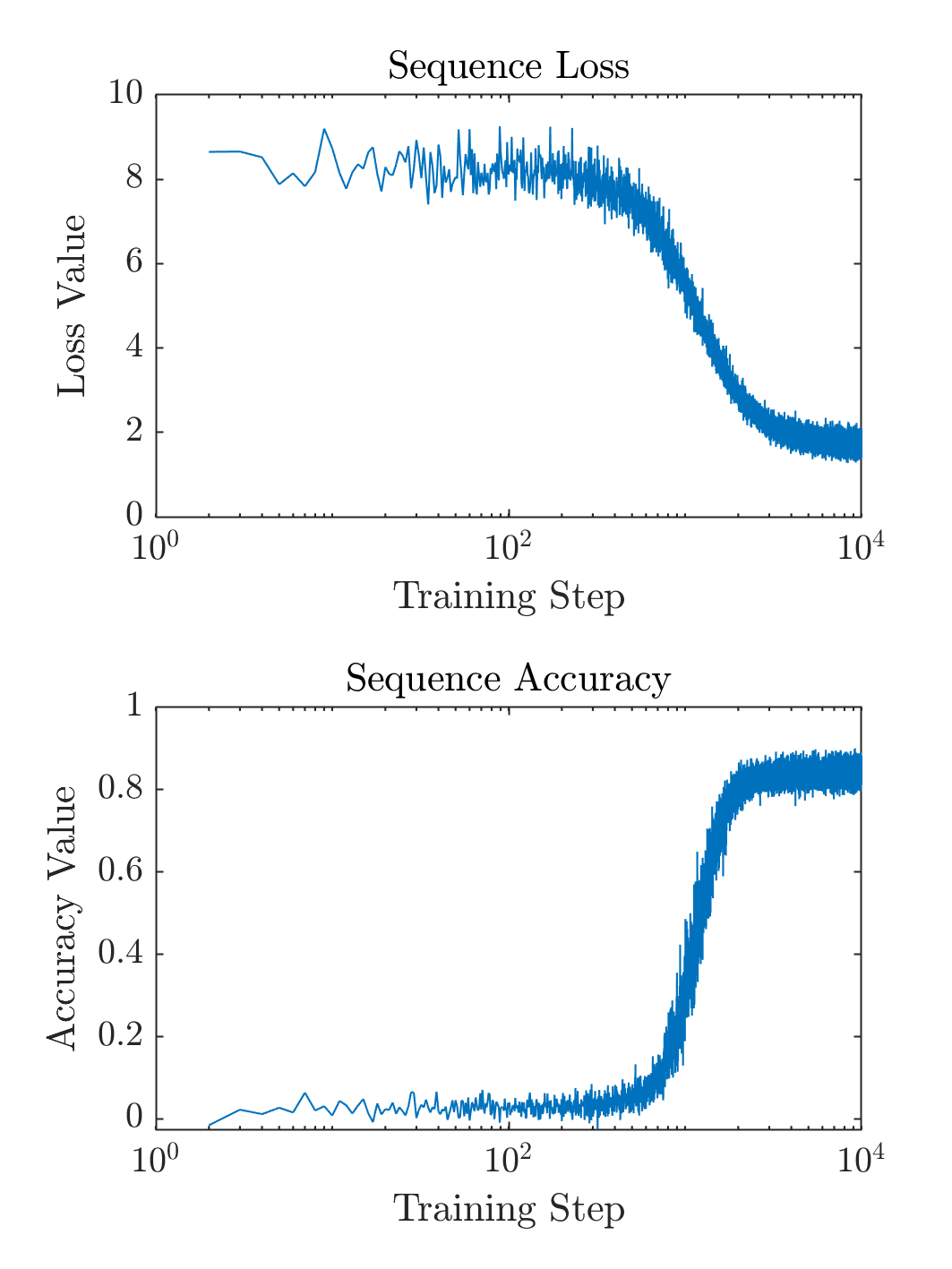}
    \caption{Loss and accuracy values during training of the best model. Note the good convergence of values to a high accuracy.}
    \label{F:LossAndAccuracy}
\end{figure}

The ML model was successfully trained and evaluated using the methods described above. The loss and accuracy values during training are shown in \cref{F:LossAndAccuracy}. Due to the low learning rate, the convergence of the training had some variability, but generally converged to a good accuracy. Convergence occurred at \num{\sim 5000} training steps. Values for the sequence loss and accuracy, as well as the standard deviation at convergence, for the best and median trained models are shown in \cref{T:Validation}.

\begin{table}[h!]
    \centering
    \begin{tabular}{|c|c|c|}
        \hline
        Validation & Best Model & Median Model \\
        \hline\hline
         Loss & 1.77 (0.16) & 1.77 (0.17) \\
         Accuracy & 0.84 (0.019) & 0.84 (0.019) \\
         \hline
    \end{tabular}
    \caption{Validation results of the training at convergence, showing the mean loss and accuracy (parenthetical values give the respective standard deviations). Note the strong stability between the best and median models.}
    \label{T:Validation}
\end{table}

\subsection{Evaluation}
\label{S:LinearResults}

\begin{figure*}[h!]
    \centering
    \includegraphics[]{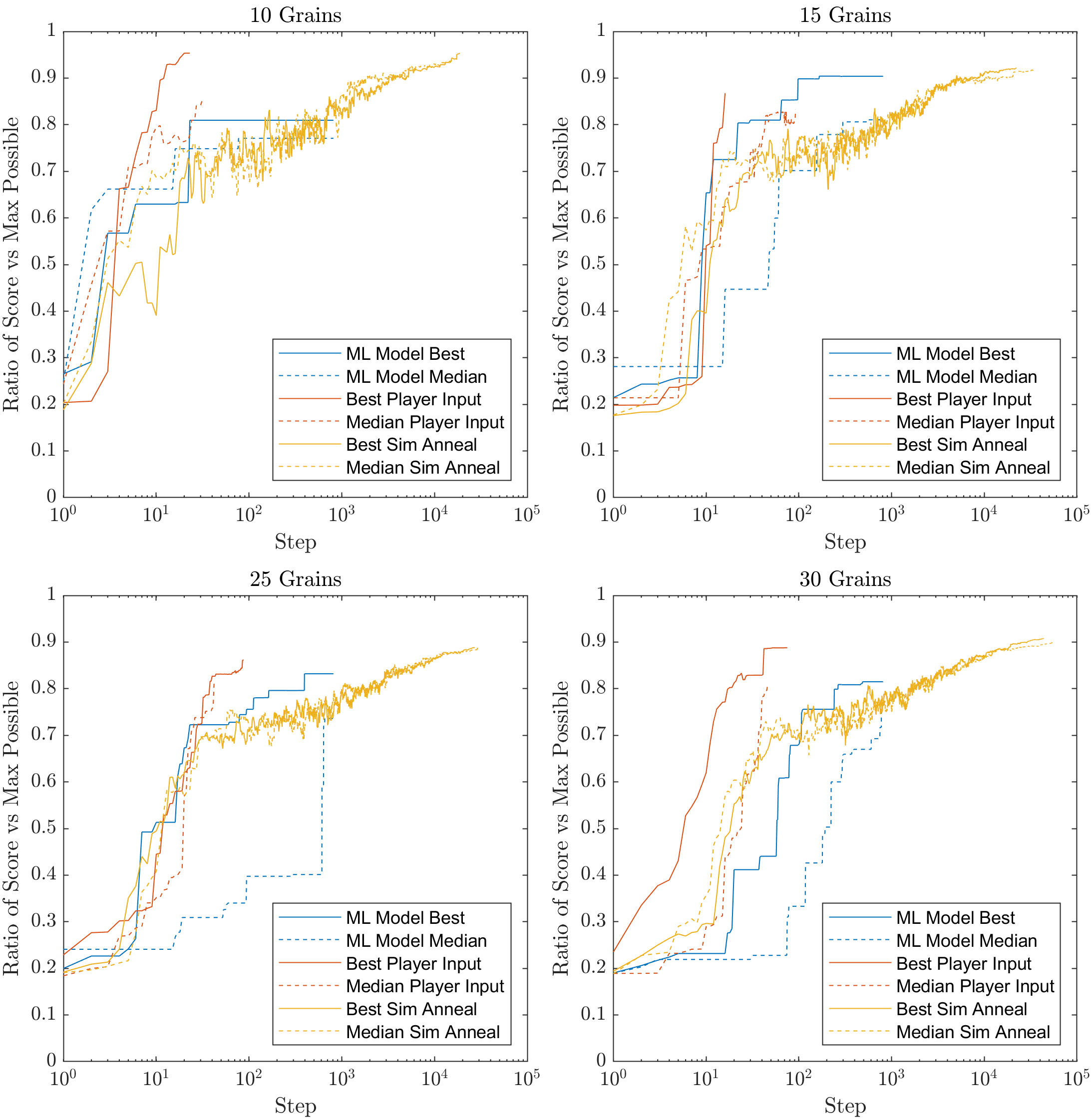}
    \caption{Time history of returns for each method (ML model, players, and SA) using the Linear constitutive model on each of the 4 evaluation microstructures. The best and median performance (by solution quality) is shown for each method.}
    \label{F:TimeHistory}
\end{figure*}

The time history of the normalized returns resulting from application of the 3 optimization strategies (players, SA, and the ML model) to each of the 4 evaluation microstructures are shown in \cref{F:TimeHistory}. Both the best and median performing trajectories (by return) are shown for each method.

\begin{table*}[h]
    \centering
    \begin{tabular}{|c|c|c|c|}
        \hline
        $n_{Grains}$ & Highest Return & $\Delta \text{Return}_{Player}$ & $\Delta \text{Steps}_{Player}$ \\
        \hline\hline
        10 & 0.810 (0.771) & -0.144 (-0.082) & -14 (-68) \\
        15 & 0.904 (0.811) & 0.037 (-0.022) & -82 (-507) \\
        25 & 0.832 (0.736) & -0.030 (-0.079) & -329 (-628) \\
        30 & 0.815 (0.771) & -0.073 (-0.037) & -465 (-736) \\
        \hline
    \end{tabular}
    \caption{Performance of the best (and median, shown parenthetically) ML model compared to the best (median) players using the Linear model as the constitutive relation for each structure of interest. $\Delta \text{Return}_{Player} = \text{Return}_{ML} - \text{Return}_{Player}$ (so negative values indicate higher player returns). $\Delta \text{Steps}_{Player} = \text{Steps}_{Player} - \text{Steps}_{ML}$ (so negative values indicate fewer steps were taken by the players to achieve an equivalent return).}
    \label{T:EvaluationPlayer}
\end{table*}
\begin{table*}[h!]
    \centering
    \begin{tabular}{|c|c|c|c|}
        \hline
        $n_{Grains}$ & Highest Return & $\Delta \text{Return}_{SA}$ & $\Delta \text{Steps}_{SA}$ \\
        \hline\hline
        10 & 0.810 (0.771) & -0.144 (-0.168) & 126 (-26) \\
        15 & 0.904 (0.811) & -0.018 (-0.107) & 4861 (266) \\
        25 & 0.832 (0.736) & -0.056 (-0.152) & 2492 (-596) \\
        30 & 0.815 (0.771) & -0.093 (-0.128) & 1427 (-86) \\
        \hline
    \end{tabular}
    \caption{Performance of the best (and median, shown parenthetically) ML model compared to the best (median) SA using the Linear model as the constitutive relation for each structure of interest. $\Delta \text{Return}_{SA} = \text{Return}_{ML} - \text{Return}_{SA}$ (so negative values indicate higher SA returns). $\Delta \text{Steps}_{SA} = \text{Steps}_{SA} - \text{Steps}_{ML}$ (so negative values indicate fewer steps were taken by SA to achieve an equivalent return).}
    \label{T:EvaluationSA}
\end{table*}

\subsubsection{ML Model vs. Players}
As intended, the ML model performance was comparable to the player performance. However, as reported in \cref{T:EvaluationPlayer}, the players achieved slightly higher returns (on average \SI[retain-explicit-plus]{+6.6}{\percent} compared to the best ML model, and \SI[retain-explicit-plus]{+6.7}{\percent} compared to the median ML model), with the exception of the 15 grain return, for which the ML model achieved a higher return. The players also required a smaller number of steps to achieve an equivalent return (the maximum ML return or the maximum player return, whichever was lower)---on average 223 fewer steps compared to the best ML model and 485 fewer steps compared to the median ML model. It is also worth noting, that while there is no obvious trend in $\Delta \text{Return}_{Player}$ with $n_{Grains}$, there does appear to be a trend of decreasing efficiency (more negative values of $\Delta \text{Steps}_{Player}$) with increasing problem size ($n_{Grains}$). Note also that while the efficiency of the ML model is lower than that of the players, the rise in returns have similar slopes (see \cref{F:TimeHistory}), meaning that the ML model can increase properties at a comparable rate to the players, though sometimes this only occurs after an initial delay.

\subsubsection{ML Model vs. SA}
When comparing the ML model to SA (see \cref{T:EvaluationSA}), we find again that, in absolute terms, the solution quality achieved by SA was slightly higher than the ML model by an amount similar to the players (i.e. the values of $\Delta \text{Return}_{SA}$ in \cref{T:EvaluationSA} are similar in magnitude to the values of $\Delta \text{Return}_{Player}$ in \cref{T:EvaluationPlayer}). However, it should be noted that SA was given a much longer time to run (the maximum number of steps allowed for SA was far greater than for players or the ML model). In contrast to the player comparison, the efficiency of the best ML model was much higher than SA (on average the best ML model required 2226 fewer steps than SA to achieve an equivalent performance). Given that the players greatly outperform SA in terms of efficiency, and the ML model is intended to emulate the players, this is one of the desired results.

\section{Discussion}

\begin{figure}[h!]
    \centering
    \includegraphics[]{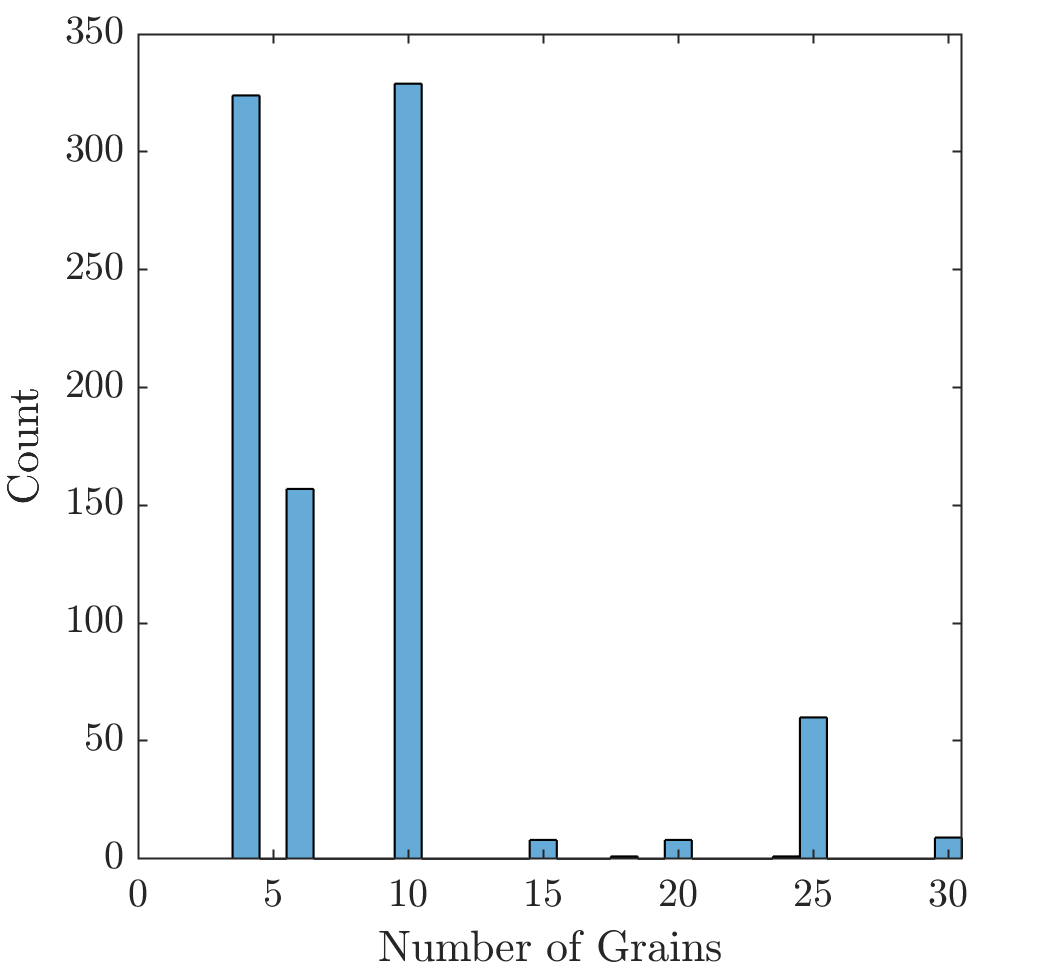}
    \caption{Distribution of problem sizes ($n_{Grains}$) in the training data for the ML model.}
    \label{F:NGrainDist}
\end{figure}

There are a number of encouraging observations about the ML model evaluation results. The first is the good solution quality match to the player data. Despite sparse training data, especially for the larger structures (see \cref{F:NGrainDist}), the ML model was capable of generating quality solutions. Second, the best ML model was capable of doing so at comparable efficiency to players, and, consequently, much greater efficiency than SA. 

\begin{table*}[h!]
    \centering
    \begin{tabular}{|c|c|c|c|}
        \hline
        $n_{Grains}$ & Highest Return & $\Delta \text{Return}_{SA}$ & $\Delta \text{Steps}_{SA}$ \\
        \hline\hline
        10 & 0.740 (0.698) & -0.046 (-0.088) & 3731 (641) \\
        15 & 0.545 (0.402) & -0.234 (-0.377) & -411 (-774) \\
        25 & 0.673 (0.512) & -0.063 (-0.104) & 3838 (-753) \\
        30 & 0.728 (0.710) & -0.047 (-0.126) & 16534 (13868) \\
        \hline
    \end{tabular}
    \caption{Performance of the best (and median, shown parenthetically) ML model compared to the best (median) SA using the Borisov/BRK model as the constitutive relation for each structure of interest. $\Delta \text{Return}_{SA} = \text{Return}_{ML} - \text{Return}_{SA}$ (so negative values indicate higher SA returns). $\Delta \text{Steps}_{SA} = \text{Steps}_{SA} - \text{Steps}_{ML}$ (so negative values indicate fewer steps were taken by SA to achieve an equivalent return).}
    \label{T:EvaluationSABRK}
\end{table*}

With these promising results, we now turn our attention to questions of generalizability of the ML model. First, as explained in \cref{S:M-ConstModel}, we investigate whether the ML model trained on data that employed one constitutive GB structure-property model (the Linear model) can be used for optimization of microstructures with a different constitutive model applied (the Borisov/BRK model), without any additional data nor any retraining.

\subsection{Generalization: Constitutive Models}
\label{S:BRKResults}

\begin{figure*}[h!]
    \centering
    \includegraphics[]{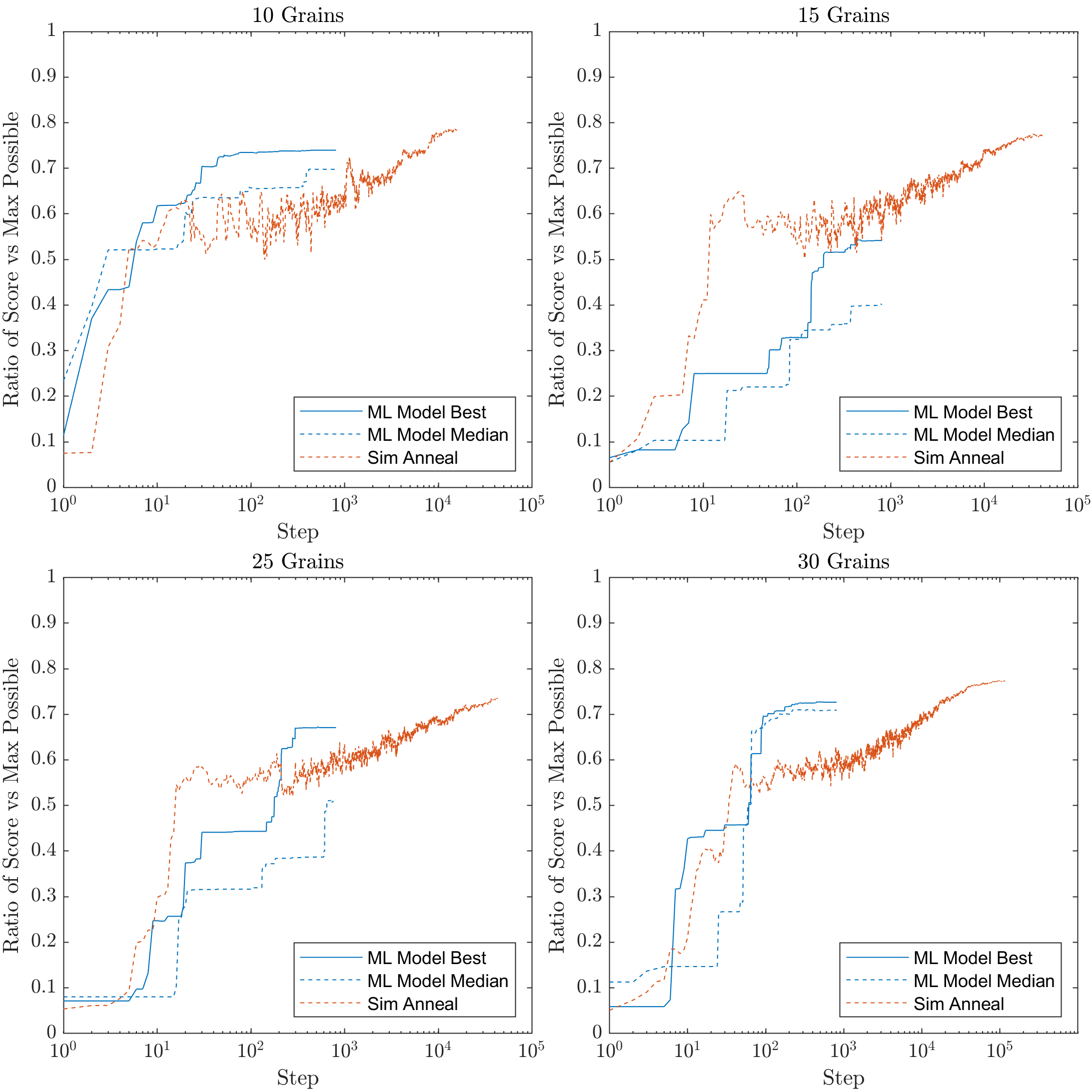}
    \caption{Time history of returns for the ML model and SA using the Borisov/BRK constitutive model on each of the 4 evaluation microstructures. The best and median performance (by solutions quality) is shown for the ML model.}
    \label{F:TimeHistoryBRK}
\end{figure*}

As described at the end of \cref{S:M-Quant}, to test generalizability of the ML model to optimization using the Borisov/BRK constitutive GB structure-property model (which it was \textit{not} trained on), we simply repeat the evaluation of the ML model for the 4 evaluation microstructures, replacing the Linear constitutive model with the Borisov/BRK model for GB property assignment. No additional training was performed. The time history of the normalized returns resulting from application of the ML model and SA (there is no player data for the Borisov/BRK model) to each of the 4 evaluation microstructures are shown in \cref{F:TimeHistoryBRK}.

At evaluation for the Borisov/BRK model, the overall returns were lower compared to when the Linear model was used, for all 4 microstructures for both ML and SA (see \cref{F:TimeHistoryBRK,T:EvaluationSABRK}). However, as reported in \cref{T:EvaluationSABRK}, the relative performance of the ML model compared to SA ($\Delta \text{Steps}_{SA}$) remained comparable to the results observed under the Linear constitutive model, in spite of not having been trained on the Borisov/BRK model (compare \cref{T:EvaluationSA,T:EvaluationSABRK}).

One notable difference, however, is that the efficiency of the ML model relative to SA is notably \textit{improved}. The best ML model required, on average, 5923 fewer steps than SA to achieve the same return (the maximum ML return or the maximum SA return, whichever was lower), and the median ML model required, on average, 3245 fewer steps than SA.

\subsubsection{Generalization: Problem Size}

The number of grains in both the training and evaluation microstructures are small compared to macroscopic polycrystals. However, one does not interrogate the entirety of macroscopic polycrystals during microstructure analysis. Rather, representative volume elements (RVEs) or sets of statistical volume elements (SVEs) are studied. An RVE is the smallest material volume whose properties (or structural characteristics) reflect those of the macroscopic material. A set of SVEs consists of smaller material volumes whose average properties (or structural characteristics) match those of the macroscopic material (thus for an SVE set to be fully defined both the size of each element \textit{and} the cardinality of the set must both be specified). Just as in microstructure \textit{analysis}, microstructure \textit{design} can also be performed in the context of an RVE or an SVE set.

Critchfield, et al. \cite{Critchfield2020}, studied RVE and SVE sizes for GBNs. The RVE size for GBNs was found to be about 200 grains \cite{Critchfield2020}, which is approximately $10\times$ the value of $n_{Grains}$ in the current training data. Fortunately, the same information can be captured by using SVEs, where 12 samples of 10 grains, or 3 samples of 50 grains, can give the same information assuming 10\% allowable error \cite{Critchfield2020}.

These SVE sizes are consistent with the training data presented here. Thus, the success of the ML model for microstructure optimization, particularly in the context of the Borisov/BRK constitutive model, suggests the potential for solving practical GBN design problems when SVEs are employed.

However, we are also interested in investigating whether the ML model can generalize to solve design problems for microstructure sizes that it has not seen in its training data. To get at this question we first consider the distribution of $n_{Grains}$ in the training data.

Due to the video game nature of the training data collection, the collected trajectories are heavily skewed towards puzzles presented earlier in the video game experience, which had fewer grains. Out of the 897 collected trajectories, only 9 trajectories were for grain numbers equal to 30, and 69 trajectories for grain number 25 or higher (see \cref{F:NGrainDist}). Thus the only evaluation microstructure whose size had an appreciable amount of corresponding training data was the 10 grain microstructure. However, in spite of the paucity of corresponding training data, the returns for the 15, 25, and 30 grain microstructures under the Linear constitutive model all \textit{exceeded} that of the 10 grain microstructure (see \cref{T:EvaluationSA}). For the Borisov/BRK constitutive model the 30 grain microstructure achieved the second highest return, and the highest efficiency gain relative to SA (see \cref{T:EvaluationSABRK}).

The goal of the position encoding, and separating actions to each individual grain, was to enable the ML model to learn correlations between GBN structure and effective properties, that would allow for generalization to microstructure sizes not included in the training data. That is, the hope was that by doing so the ML model would ``understand'' the relationship between GBN structure and properties in some kind of generalizable way. The success of the ML model for the 15, 25, and 30 grain evaluation microstructures, for which very little training data existed, provides some evidence for this kind of learning and suggests the potential of such size generalizability.

To investigate this further, we performed an experiment in which we retrained the same ML model for the Linear constitutive model player data, but removed subsets of training data. We then evaluated the resulting ML models using the Borisov/BRK constitutive GB structure-property model. This is a particularly challenging test in that it involves a systematic reduction of the training data \textit{and} testing against a GB structure-property model that the ML model was not trained on. 

\begin{figure}
    \centering
    \includegraphics[]{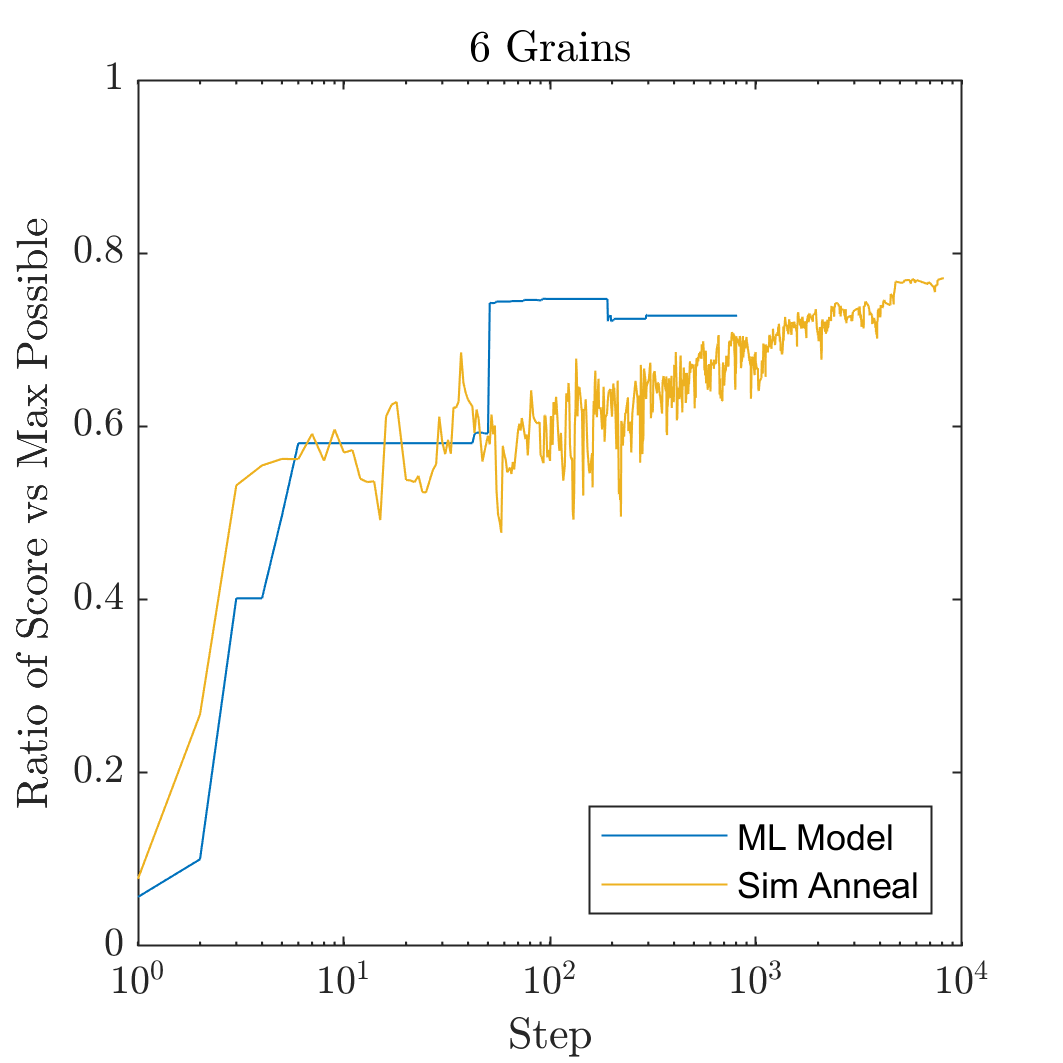}
    \caption{Trajectories of a model that is only trained on player data containing microstructures with 10 or more grains. Note the comparable performance to the 10 grain problem in \cref{F:TimeHistoryBRK}, which used all of the training data.}
    \label{F:SmallGrainCompare}
\end{figure}

The first retraining removed all training data for grain numbers less than 10, and a second retraining removed all grain numbers greater than 10. This removed 481 trajectories (53\%) from our training set when removing the small grain number trajectories, and 87 trajectories (9\%) when removing the large trajectories. After performing the stated retraining, the model which removed the small $n_{Grains}$ trajectories converged to a sequence accuracy of 0.908 (SD=0.007). This is noticeably better performance than before, but could be attributed to overfitting to a smaller dataset (more than half the training data was removed). However, when the trained model was evaluated using the Borisov/BRK constitutive model on a 6 grain sample (fewer grains than any of the supplied training data), we see that the performance is comparable to the model trained on the whole dataset when evaluated on the 10 grain evaluation microstructure (compare \cref{F:SmallGrainCompare} to \cref{F:TimeHistoryBRK,T:EvaluationSABRK}). 

\begin{figure}
    \centering
    \includegraphics[]{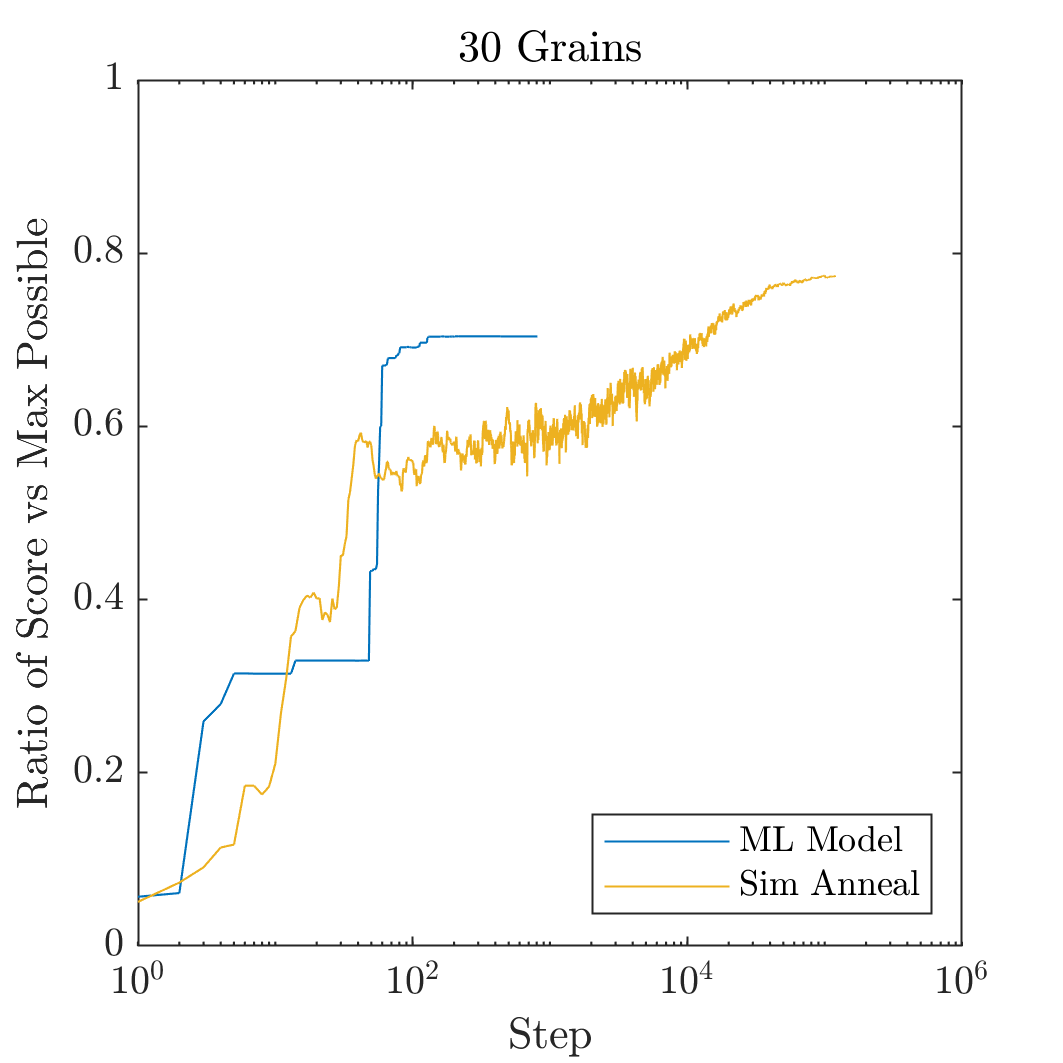}
    \caption{Trajectories of a model that is only trained on player data containing microstructures with 10 or fewer grains. Note the similar performance to the 30 grain case in \cref{F:TimeHistoryBRK} even though no training data was given for this grain number.}
    \label{F:SmallToLargeCompare}
\end{figure}

When removing the large $n_{Grains}$ trajectories, the training converged to a sequence accuracy of 0.832 (SD=0.019), which is close to the accuracy of the fully trained model. This can be expected because very little data is removed in this different training case. When this model is applied to the 30 grain sample (much larger than the training set) using the Bulatov/BRK model it compares very closely to the previously trained model (compare \cref{F:SmallToLargeCompare} to \cref{F:TimeHistoryBRK,T:EvaluationSABRK}).

Thus we find that the ML model is successful and shows comparable performance even when applied to evaluation microstructures whose size falls outside of the range of the training data.

\subsection{Interpretability of the Transformer}

\begin{figure}[ht]
    \centering
    \includegraphics[]{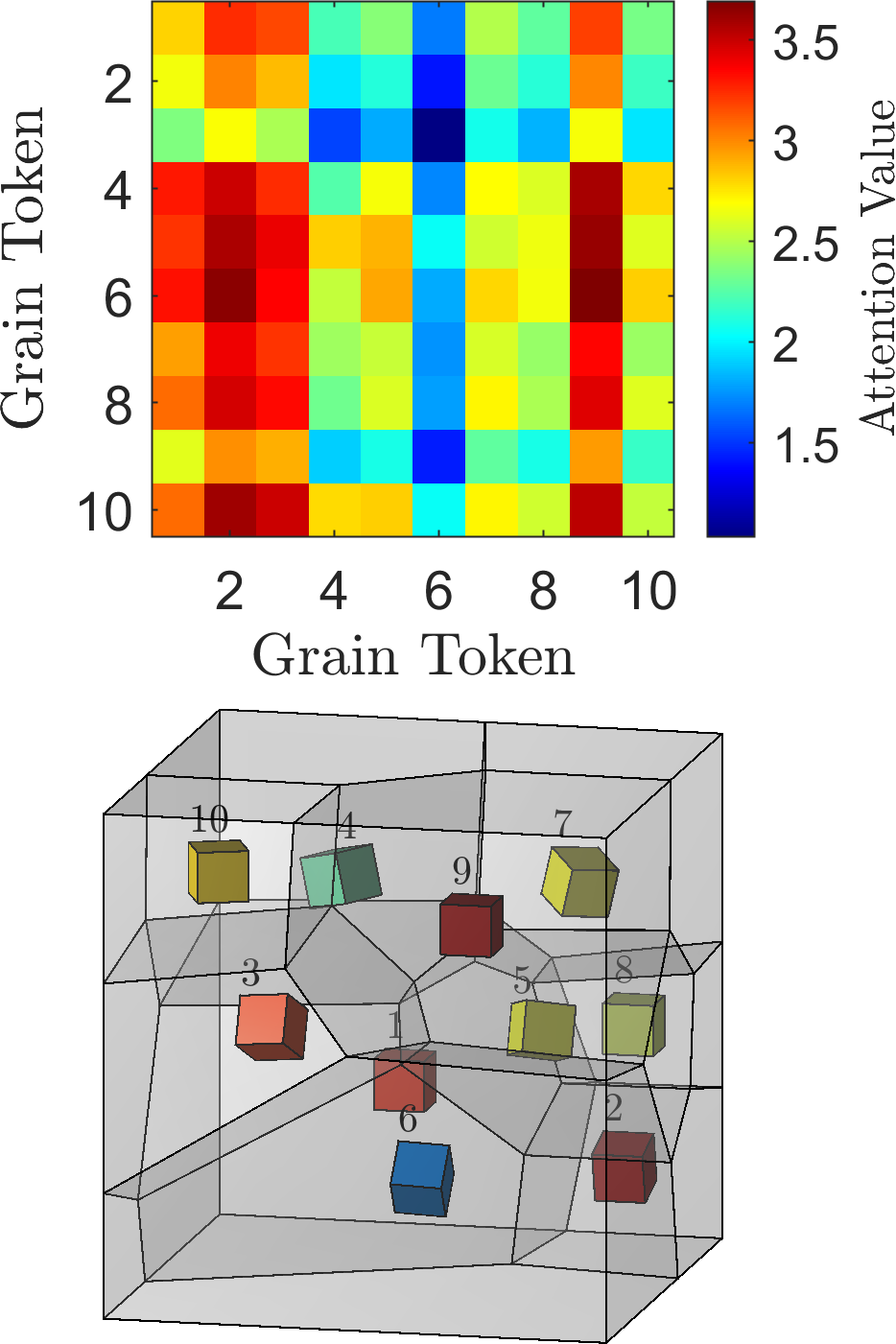}
    \caption{Visualization of the attention block of the state at time step 51, as well as a projection of the attention scores of the chosen grain (grain 4) onto the microstructure.}
    \label{F:AttentionGridToMicro}
\end{figure}

As a final discussion point, we consider the potential for using the transformer attention scores to interpret/understand the ML model's chosen actions. The attention scores, $Q \times K^{\mathsf{T}}$ from \cref{E:Transformer}, provided by the transformer layer in the model have been used previously to visualize what state inputs the transformer focuses on to choose a given action. The authors of the original Decision Transformer \cite{Chen} and those of the Multi-Game Transformer \cite{Lee2022} claim this attention matrix can visualize where the model is ``looking'' when making decisions. 

An example of the attention scores for the GBN ML model for a given state step and the chosen action can be seen in the heatmap shown in \cref{F:AttentionGridToMicro}. Each row represents the attention scores of the corresponding grain (agent) ``looking'' at itself and the other grains in the simulation. \cref{F:AttentionGridToMicro} represents step 51 of this optimization trajectory (where a large property increase resulted). At this step, the ML model chose to apply a manual rotation to grain 4 (corresponding to row 4 of the attention heatmap). 

It is interesting to note that grains 2 and 9 have the highest attention scores (where the transformer is ``looking''), and grain 2 is not a nearest neighbor to the selected grain. The fact that non-nearest neighbors to the selected grain have the highest attention scores could reflect that optimal GBN properties are dependent on intermediate to long range connectivity effects, and suggest that the ML model has learned something about such longer-range relationships.

We note, however, that interpretation of attention scores is still debated in the ML community as the attention shown is simply the cross product of the final two dimensions, $T\text{ and } h$, of two Linear layers, and robust definitions and understanding are still being developed \cite{Achache,Chefer2021}.

\section{Conclusions}

Optimization of GBNs is difficult due to their high-dimensional state space, but would enable design of materials with enhanced properties in multiple application areas. Prior work \cite{Adair2022} showed that leveraging human spatial reasoning and intuited dimensionality reducing heuristics via a video game was a viable route for obtaining optimization pathways, but such data are expensive and slow to obtain.

In the present work, we demonstrated how such human optimization trajectories for GBNs can be used to train a Decision Transformer ML model to perform microstructure design. The resulting ML model uses all crystallographic information and long range connectivity to generate microstructures with optimized macroscopic properties for a given constitutive GB structure-property model. This implementation of the Decision Transformer expands the capabilities of the model to allow multiple concurrent actions during a decision step, rather than a single action per set of observations like previous implementations. This allowed the model to act on a variable amount of grains with a set of actions, and used the same position encoding as the observation space.

When compared to simulated annealing (a popular global optimization algorithm), we found that the ML model obtained solutions that were of comparable quality (on average \SI{92}{\percent} as optimal as the SA solutions). However, we found that the ML model was far more efficient, requiring three orders of magnitude fewer iterations to do so.

We found that the ML model was capable of generalizing to constitutive structure-property models on which it was not trained. In particular, we trained the ML model on human player data from design problems that employed a simple (and computationally inexpensive) GB constitutive structure-property model. We then used the trained model to solve design problems for a high-fidelity (and computationally expensive) GB structure property model \textit{without} any additional training. We found that the ML model showed similar and in some cases even better performance (relative returns and efficiency gains over SA). This capability is particularly useful when training data for the constitutive structure-property model of interest is unavailable, but one does have access to training data for a different (e.g. simpler) structure-property model.

We demonstrated that the ML model can solve problem sizes relevant to real microstructure design applications, particularly when statistical volume elements (SVEs) are employed. We also found evidence for generalizability to problem sizes outside of the range considered in training data, which may allow for solution of larger design problems.

Finally, we gave a brief illustration of how the underlying attention scores can be visualized, which may possibly lead to interpretation of the optimization strategies employed by the ML model.

Based on these results, we believe Decision Transformer based models may allow for efficient design of GBNs, and perhaps materials design problems generally.

\section*{Acknowledgement}
\label{S:acknowledgement}

The material presented here is based upon work supported by the National Science Foundation under Grant No. DMR-1654700.

\section*{Declaration of Interest}
\label{declaration}

The authors declare that they have no known competing financial interests or personal relationships that could have appeared to influence the work reported in this paper.

\section*{Data Availability Statement}
\label{data}

Both data and model code are available in the linked repository \cite{Adair_A-Decision-Transformer-Approach-to-Grain-Boundary-Network-Optimization_2024}.
\bibliographystyle{elsarticle-num-names}
\bibliography{paper3References.bib}
\end{document}